\documentclass[11pt]{article}
\usepackage{aaspp4}
\newcommand\fsrc{\mbox{G359.87$+$0.18}} 
\newcommand\sgra{\mbox{Sgr~A${}^*$}}
\newcommand\mjybm{\mbox{mJy~beam${}^{-1}$}}
\newcommand\kms{\mbox{km~s${}^{-1}$}}
\newcommand\delgc{\Delta_{\mathrm{GC}}}
\newcommand\dgc{D_{\mathrm{GC}}}

\newcommand\aips{\textsc{aips}}

\received{1998 August~12}
\revised{1998 November~10}
\accepted{1998 November~18}
\journalid{515}{1999 April~10}
\paperid{39028}

\lefthead{Lazio et al.}
\righthead{\fsrc}

\begin{document}
\title{\fsrc: An FR~II Radio Galaxy 15~Arcminutes from \sgra.\\
	Implications for the Scattering Region in the Galactic Center}

\author{T.~Joseph~W.~Lazio\altaffilmark{1}}
\affil{NRL, Code~7210, Washington, DC, 20375-5351;
	lazio@rsd.nrl.navy.mil}
\altaffiltext{1}{NRC-NRL Research Associate}

\author{K.~R.~Anantharamaiah\altaffilmark{2} and W.~M.~Goss}
\affil{NRAO, P.~O.\ Box~0, Socorro, NM 87801; anantha@aoc.nrao.edu;
	mgoss@nrao.edu}
\altaffiltext{2}{On leave from Raman Research Institute, Bangalore
	560~080, India}

\author{Namir E.~Kassim}
\affil{NRL, Code~7213, Washington, DC, 20375-5351;
	nkassim@rsd.nrl.navy.mil}

\and
\author{James M.~Cordes}
\affil{Department of Astronomy, Cornell University and NAIC, 520 Space
	Sciences Bldg., Ithaca, NY, 14853-6801;
	cordes@spacenet.tn.cornell.edu}

\authoraddr{Address for correspondence:
	T. Joseph W. Lazio
	NRL, Code 7210
	Washington, DC  20375-5351}	

\begin{abstract}
\fsrc\ is an enigmatic object located 15\arcmin\ from \sgra.  It has
been variously classified as an extragalactic source, Galactic jet
source, and young supernova remnant.  We present new observations of
\fsrc\ between~0.33 and~15~GHz and an \ion{H}{1} absorption spectrum
and use these to argue that this source is an Faranoff-Riley~II radio
galaxy.  We are able to place a crude limit on its redshift of $z
\gtrsim 0.1$.  The source has a spectral index~$\alpha < -1$ ($S \propto \nu^\alpha$), suggestive of a radio galaxy with a
redshift $z \gtrsim 2$.

The scattering diameters of \sgra\ and several nearby OH masers
($\approx 1\arcsec$ at 1~GHz) indicate that a region of enhanced
scattering is along the line of sight to the Galactic center.  If the
region covers the Galactic center uniformly, the implied diameter for
a background source is at least 600\arcsec\ at~0.33~GHz, in contrast
with the observed 20\arcsec\ diameter of \fsrc.  Using the scattering
diameter of a nearby OH maser OH~359.762$+$0.120 and the widths of
two, nearby, non-thermal threads, G0.08$+$0.15 and G359.79$+$0.17, we
show that a uniform scattering region should cover \fsrc.  We
therefore conclude that the Galactic center scattering region is
inhomogeneous on a scale of~5\arcmin\ ($\approx 10$~pc at a distance
of 8.5~kpc).  This scale is comparable to the size scale of molecular
clouds in the Galactic center.  The close agreement between these two
lengths scales is an indication that the scattering region is linked
intimately to the Galactic center molecular clouds.
\end{abstract}


\section{Introduction}\label{sec:intro}

The observed diameter of \sgra, the compact source in the Galactic
center (GC), scales as $\lambda^2$, as expected if interstellar
scattering from microstructure in the electron density determines the
observed diameter (Davies, Walsh, \& Booth~1976).  The observed
diameter of \sgra\ is now known to scale as $\lambda^2$ from~30~cm
to~3~mm and to be anisotropic at least over the wavelength range 21~cm
to~7~mm (\cite{bzkrml93}; \cite{krichbaumetal93}; \cite{rogersetal94};
\cite{y-zcwmr94}; \cite{bb98}).  Maser spots in OH/IR stars
within~25\arcmin\ of \sgra\ also show enhanced, anisotropic angular
broadening (\cite{vfcd92}; \cite{fdcv94}).  These observations
indicate that a region of enhanced scattering with an angular extent
of at least 25\arcmin\ in radius (60~pc at 8.5~kpc) is along the line
of sight to the Galactic center.  Assuming that the region covers the
Galactic center uniformly, the scattering diameter of an extragalactic
source seen through this region should be larger than those of
Galactic sources by the ratio~$\dgc/\delgc$, where $\dgc$ is the
Sun-GC distance and $\delgc$ is the GC-scattering region separation
(\cite{vfcd92}).  Lazio \& Cordes~(1998a, b) constrain this ratio to
be $\dgc/\delgc \approx 50$, so that an extragalactic source seen
through this scattering region should have a diameter of at least
75\arcsec\ at~1~GHz.

In their analysis, Lazio \& Cordes~(1998a, b) paid particular
attention to the source~\fsrc.  First identified by Isaacman~(1981,
source 35W44) and Yusef-Zadeh \& Morris~(1987, source~J), \fsrc\ was
initially (and tentatively) classified as extragalactic by
Anantharamaiah et al.~(1991, p.~280).  Their identification of this
source as extragalactic was based on its compact morphology and flux
density of~0.58~Jy at~0.33~GHz.  This extragalactic classification was
further strengthened by~8.5 and~15~GHz observations by W.~M.~Goss \&
K.~R.~Anantharamaiah~(1990, unpublished) showing the source to have a
Fanaroff-Riley~II radio galaxy morphology with a 4\arcsec\ separation
between components.

Based on~0.33 and~1.4~GHz images of the source, Lazio \&
Cordes~(1998a) pointed out that the source could not be extragalactic
\emph{and} affected by the hyperstrong scattering region responsible
for the enhanced angular broadening of \sgra\ and the OH masers.  They
suggested that the source might be an X-ray--quiet version of a
Galactic jet source, like 1E1740.7$-$2942 (\cite{mrcpl92}).
Yusef-Zadeh, Cotton, \& Reynolds~(1998a) have recently considered a
number of classifications for the source including young supernova
remnant, radio supernova, or nova remnant, based on high-resolution,
total intensity images at~15~GHz and polarized intensity images
at~5~GHz.  They did not make an unambiguous classification, but
favored a young supernova remnant as most likely.

This paper reports new high resolution images of \fsrc\ between~0.33
and~15~GHz and an \ion{H}{1} absorption spectrum.  We shall conclude
that the source is most likely to be an extragalactic source.  In
\S\ref{sec:observe} we present the new observations.  In
\S\ref{sec:id} we summarize why we believe that \fsrc\ is likely to be
extragalactic, \S\ref{sec:scatter} discusses what this implies for the
scattering region in the \hbox{GC}, and in \S\ref{sec:conclude} we
present our conclusions.  Throughout we assume that the Galactic
center is at distance of~8.5~kpc, implying that $1\arcmin = 2.5$~pc.
We also use the convention that the spectral index~$\alpha$ is given
by $S \propto \nu^\alpha$.

\section{Summary of Observations}\label{sec:observe}

Our observations of the source \fsrc\ span 0.33 to~15~GHz.  Some of
the observations are new, while in other cases, we have re-analyzed
existing data to obtain higher resolutions or dynamic ranges.
Table~\ref{tab:observe} is our observing log.  A summary of
lower-resolution observations of \fsrc\ is contained in Lazio \&
Cordes~(1998a).

Image analysis was within \aips\ using standard reduction procedures,
except where noted.  We discuss each frequency separately.

\subsection{0.33~GHz}\label{sec:Pband}

As part of a search for radio pulsars in the GC, we observed \sgra\
at~0.33~GHz with the VLA in the A configuration.  The radio pulsar
survey will be described elsewhere (\cite{lcka98}).  The FWHM primary
beam of the VLA at~0.33~GHz is 2\fdg5, large enough that little
primary beam attenuation occurs over the 15\arcmin\ distance between
\sgra\ and \fsrc.  However, at~0.33~GHz the VLA can no longer be
considered co-planar, and the standard assumption of a two-dimensional
Fourier transform relation between the measured visibilities and the
sky brightness distribution is no longer valid (\cite{cp92}).

We imaged \fsrc\ using the program \texttt{dragon} within NRAO's
Software Development Environment.  Rather than attempt to image the
entire primary beam, \texttt{dragon} allows small portions of the sky
to be imaged (\cite{cp92}).  For each small image patch
\texttt{dragon} inserts an appropriate phase shift, thereby restoring
the validity of the assumption of co-planar baselines.
Figure~\ref{fig:Pband} shows the image patch containing \fsrc\ and the
thread G359.79$+$0.17 to the south.

At~0.33~GHz \fsrc\ appears as a single, resolved component.  Below we
shall identify the appearance at~0.33~GHz as due partially to
scattering.

The integrated flux density for \fsrc\ is 0.24~Jy.  This value is only
40\% of that determined by Anantharamaiah et al.~(1991, 0.58~Jy) from
VLA observations.  This difference arises because we use data solely
from the A configuration while they used data from both A- and B
configurations.  The B configuration is more compact than the A
configuration, so the source is not as strongly resolved; the flux
density reported by Anantharamaiah et al.~(1991) is therefore more
accurate.

\subsection{1.5~GHz}\label{sec:Lband}

We observed \fsrc\ with the VLA at~1.42 and~1.66~GHz.  The data from
the two frequencies were calibrated separately and combined before
imaging.  Figure~\ref{fig:Lband} shows the resulting image.

\fsrc\ consists of three, nearly collinear components.  In keeping
with the nomenclature of Yusef-Zadeh et al.~(1998a), the stronger
component to the northeast is component~A, the central component is
component~B, and the faint, previously-unrecognized component to the
southwest is \hbox{component~C}.  While nearly collinear, the source
is asymmetric, with component~C at a larger distance from component~B
than is \hbox{component~A}.  Component~C is also resolved with an
extension pointing toward the two other components, while components~A
and~B are, at best, only slightly resolved.  The apparent connection
between components~A and~B is an artifact of the beam, $2\farcs5
\times 1\arcsec$.

\subsection{8.5~GHz}\label{sec:Xband}

We observed \fsrc\ for 105~min.\ on 1998 March~11 and April~11 with
the \hbox{VLA}.  All three components were detected.  Furthermore, in
the images from both days, component~B showed a faint, jet-like
feature pointing toward \hbox{component~A}.  There was no variation
either in the structure or in the flux density of the components
between 1998 March~11 and April~11.  Figure~\ref{fig:Xband}a shows the
high-resolution image of \fsrc\ formed by combining the observations
from these two epochs.

We also reprocessed a lower resolution, 30-min.\ observation of \fsrc\
acquired with the VLA on 1990 October~12.  Figure~\ref{fig:Xband}b
shows the image formed by combining all data from the 1990 and~1998
observations.  The lower resolution data are more sensitive to
extended structure, and component~C is now more easily detected.

\subsection{15~GHz}\label{sec:Uband}

We acquired a total of 7~hr of observation on \fsrc\ in 1998~April
with the \hbox{VLA}.  The images made from the individual observations
show only components~A and~B.  There is no variation in the structure
on the three days.  We also reprocessed a lower resolution, 30-min.\
observation of \fsrc\ acquired with the VLA on 1990 October~12.

Figure~\ref{fig:Uband} shows the image produced from the combined 1990
and~1998 data.  Component~A is resolved into a two sub-components,
with one of the two being edge-brightened.  The line joining the two
sub-components is at an angle of~70\arcdeg\ to the axis of the
large-scale morphology.  Component~B remains unresolved, but the jet-like
feature is no longer present, indicating that it has a steep spectrum.
We find a spectral index $\alpha < -2.8$, though with considerable
uncertainty ($\approx 20$--30\%).  Our estimate for the flux density
of the jet at~8.5~GHz may be contaminated by flux from the central
part of component~B, which would make the spectrum of the jet appear
steeper than it actually is.

\subsection{\ion{H}{1} Absorption Spectrum}\label{sec:hi}

We re-analyzed VLA archive \ion{H}{1} absorption spectra made toward
\sgra\ (Plante, Lo, \& Crutcher~1991).  \fsrc\ is sufficiently close
to \sgra\ that the primary beam attenuation is approximately 50\%, and
the delay beam attenuation is not significant.  Thus, we were able to
obtain a spectrum for both \fsrc\ and \sgra.  The spectra toward these
two sources are shown in Figure~\ref{fig:hi}.

The $-50$ (``3-kpc arm'') and $+40$~\kms\ features are seen in both
spectra.  Absorption is also seen at~0~\kms\ toward both sources, but
the width of the absorption line is about twice as wide toward \fsrc\
as it is toward \sgra\ (40~\kms\ vs.\ 20~\kms).  The large width of
the 0~\kms\ gas towards \fsrc\ is similar to that seen toward
B1739$-$298 ($\ell = 358\fdg92$, $b=0\fdg07$), an extragalactic source
approximately 1\arcdeg\ away from \sgra\ (\cite{dkvgh83}).  In both
cases the large width of the gas at~0~\kms\ may be a blend of two
components---one centered near~0~\kms, the other near~$-20$~\kms.  The
gas at~$-20$~\kms\ does not appear in the spectrum of \sgra\ and
presumably lies beyond \sgra.

\section{Identification of \fsrc\ as an FR~II Radio
	Galaxy}\label{sec:id}

We classify \fsrc\ as extragalactic, most likely an FR~II radio
galaxy.  Our classification is motivated by four criteria: morphology,
spectrum, polarization, and lack of structural or flux density
variation.  We have also been able to place crude limits on the
distance to the source, which are consistent with it being
extragalactic.

\subsection{Morphology}\label{sec:morph}

The source clearly shows the morphology of an FR~II source.  Our 1.5
and~8.5~GHz images (Figures~\ref{fig:Lband} and~\ref{fig:Xband}b) show
two components (A and~C) surrounding a central component (B), which we
identify as the core of the source.  Our highest resolution, 15~GHz
image (Figure~\ref{fig:Uband}) resolves component~A into two
sub-components, one of which is possibly an edge-brightened hot spot.
At~8.5~GHz, component~C is elongated in the direction toward
component~B, and component~B has a steep-spectrum, jet-like feature
pointed toward \hbox{component~A} (Figure~\ref{fig:Xband}a).  The
source is asymmetric; the separation between components~A and~B is
4\arcsec, and the separation between components~B and~C is 9\arcsec.
This asymmetry could reflect a number of effects: the orientation of
the source with respect to the line of sight, inhomogeneities in the
ambient medium, and precession of the jet.  A denser medium
surrounding component~A might also explain why it is considerably
brighter than \hbox{component~C}.

Within component~A the line joining the two hot spots is oriented at
an angle of approximately 70\arcdeg\ to the line joining components~A
\hbox{and~B}.  Such misalignment is a common feature of FR~II radio
galaxies and is thought to result from changes in the jet orientation
(\cite{cb96}).

\subsection{Spectrum}\label{sec:spectrum}

At frequencies higher than 1.5~GHz, all components of \fsrc\ have
extremely steep spectra, $\alpha \lesssim -1$.
Figure~\ref{fig:spectra} shows the spectra of the individual
components and of the entire source, as derived from our measurements,
and Table~\ref{tab:fluxdensity} tabulates the flux density of the
components.  Over the frequency range 1.5--15~GHz, component~A has a
spectral index of $-1.1$, component~B has a spectral index of $-0.7$,
and, between~1.5 and~8.5~GHz, component~C has a spectral index of
$-1.1$.  At~15~GHz we can place only an upper limit ($5\sigma$)
of~0.24~mJy on the flux density of component~C, which is slightly
lower than that expected by an extrapolation of its spectrum, namely
0.3~mJy.  Using Anantharamaiah et al.'s~(1991) value for the 0.33~GHz
flux density of~0.58~Jy and assuming that all three components
contribute to the flux at~0.33~GHz, the spectrum flattens between~0.33
and~1.5~GHz with $\alpha \approx -0.8$.

Though our flux density measurements at different frequencies were
obtained from observations with differing resolutions---with the
exception of the 0.33~GHz observations (\S\ref{sec:Pband})---the
components are sufficiently compact that this should introduce little
error.  In order to verify that this is the case, we have compared
flux density measurements made at~8.5~GHz, determined both with and
without the 1990 observations.  The lower resolution 1990 observations
should be more sensitive to extended flux that might be missed in our
higher resolution 1998 observations.  We find that the flux density
measurements, determined with and without the 1990 observations, agree
to within 10\% for both components~A \hbox{and~B}.

Yusef-Zadeh et al.~(1998a) used matched resolution, nearly simultaneous
observations to find $\alpha_{\mathrm{A}} = -1.6$ and
$\alpha_{\mathrm{B}} = -1.3$ for components~A and~B, respectively,
between~5 and~15~GHz.  Lazio \& Cordes~(1998a) find
$\alpha_{\mathrm{A}} \approx -1$ for component~A over the frequency
range 0.33--15~GHz, though they used flux density measurements made at
differing resolutions at different epochs and assumed that component~A
was responsible for all of the flux at~0.33~GHz (viz.\
\S\ref{sec:image}).

Radio sources with $\alpha < -1$ belong to the class of ultra-steep
spectrum (USS) sources (\cite{rlmcs94}).  Members of this class
include distant radio galaxies, head-tail galaxies, galaxy cluster
radio halos, ``fading'' radio galaxies, and pulsars.  Of these, \fsrc\
is most likely to be a distant radio galaxy; it does not have the
morphology of a head-tail galaxy, both radio halos and fading radio
galaxies are quite rare, and the lack of variation (see below)
indicates that it is probably not Galactic.  In addition to its
spectrum, \fsrc\ has two other characteristics typical of USS, distant
radio galaxies (\cite{rlmcs94}): First, the majority of USS sources
consist of more than one component, with 10\% being triples.  Second,
the 12\arcsec\ separation between components~A and~C is comparable to
the median angular size of USS radio sources---for USS radio sources
having a 0.33~GHz flux density of 0.5~Jy, the median angular size
at~1.4~GHz is approximately 10\arcsec.

An extrapolation of the high-frequency spectrum of \fsrc\ to~0.33~GHz
predicts a flux density of~0.8~Jy, in contrast to the observed value
of of~0.58~Jy (\cite{apeg91}).  We can identify both intrinsic and
extrinsic causes for such a high-frequency steepening of the source's
spectrum.  The intrinsic cause would be a depletion of the high-energy
electrons responsible for the synchrotron emission.  This depletion
could result from effects such as inverse Compton losses from the
radiation field of the central engine (e.g., \cite{bl95}) or from a
higher cosmic background radiation field at earlier epochs (e.g.,
\cite{crmph98}).  The extrinsic cause would be free-free absorption in
the \hbox{GC}.  An optical depth of $\tau \approx 0.3$ at~0.33~GHz
would be sufficient to account for this difference.  Optical depths
$\tau \gtrsim 1$ at~0.33~GHz are seen elsewhere in the GC, e.g., near
the radio arch (\cite{apeg91}), and an extended, low-density
\ion{H}{2} region covers the inner degree or so of the GC
(\cite{mp79}; \cite{ara97}; see also \S\ref{sec:imply}).  From our
limited spectral coverage of the source's spectrum, however, we cannot
distinguish between these various possibilities.

\subsection{Polarization}\label{sec:polarize}

Yusef-Zadeh et al.~(1998a) show that the edge-brightened areas in
component~A are significantly polarized (6--16\%) at~5~GHz.  This
level of polarization is consistent with that seen in the lobes of
FR~II radio galaxies. 

Because of large Faraday rotation, the intrinsic magnetic field
direction within component~A cannot be determined.  From two closely
spaced frequencies near~5~GHz, Yusef-Zadeh et al.~(1998a) derive a
Faraday rotation measure of~3000~rad~m${}^{-2}$.  Comparable Faraday
rotation measures are seen along other lines of sight to the GC
(\cite{ittkt84}; \cite{tihtk85}; \cite{y-zm87}; \cite{gnec95};
Yusef-Zadeh, Wardle, \& Parastaran~1997), and some FR~II radio
galaxies are embedded within a magnetized intracluster medium which
could contribute to additional Faraday rotation (\cite{cb96}).

\subsection{Variability}\label{sec:flux}

Our observations, combined with those summarized in Lazio \&
Cordes~(1998a), provide a 10--15~yr time span over which \fsrc\ has
been observed between~0.33 and~15~GHz.  On short time scales (between
a few days to a month) the source shows no structural variations, and
the flux density is essentially constant.  The angular resolution of
many of the older observations (\cite{lc98a}) was not as high as in
the present observations, so that a detailed comparison of the
structure of the source at different epochs is not possible.
Components~A and~B have been detected, at approximately the same
angular separation from each other, since 1986.  Flux density
comparisons are also hampered by the differing resolutions at the
various different epochs.  Nonetheless, the flux density of the source
does not appear to have varied by more than 10\% at any frequency over
this this time span, and flux density variations of roughly 10\% may
be the result of (extrinsic) refractive scintillation (Rickett, Coles,
\& Bourgois~1984).

\subsection{Distance}\label{sec:distance}

We can place only crude limits on the distance to \fsrc, but these are
consistent with it being extragalactic.  The \ion{H}{1} spectra of
\fsrc\ and B1739$-$298 are similar: Both are dominated by an
absorption feature centered approximately on~0~\kms, with a
considerable width, 40~\kms, and optical depth, $\tau \ge 2$.  In both
sources there is a suggestion that this large width is the
superposition of two velocity components, one centered on 0~\kms, the
other at~$-20$~\kms.  The 20~\kms\ gas does not appear in the spectrum
of \sgra\ and presumably lies beyond it.  The forbidden velocities
near the GC make determining a distance from a rotation curve
problematic, but the spectrum of \fsrc\ indicates that it is at least
as far away as the GC and is consistent with an extragalactic
distance.

We can also place a lower limit on the distance to \fsrc\ by requiring
that its power output be consistent with that of other FR~II galaxies.
Extrapolating the (high-frequency) spectrum of \fsrc\ to~0.178~GHz, we
find that its flux density would be 1.7~Jy.  In order for its
luminosity to exceed the break between FR~I and FR~II galaxies at $2
\times 10^{25}$~W~Hz${}^{-1}$~sr${}^{-1}$, \fsrc\ must be at a
distance $D > 300$~Mpc ($z > 0.1$ for $H = 100$~\kms~Mpc${}^{-1}$ and
$q = 1/2$).

By comparision with other USS distant radio galaxies, a redshift of $z
\gtrsim 2$ is indicated (\cite{rlmcs94}).  In a sample of
high-redshift radio galaxies selected on the basis of $z > 2$, Athreya
et al.~(1997) find that most of them have unresolved cores with
spectra $\alpha < -0.5$, similar to that of \hbox{component~B}.

\section{Scattering Toward the Galactic Center}\label{sec:scatter}

\subsection{Diameter and Shape of \fsrc}\label{sec:image}

Figure~\ref{fig:overlay} shows a 8.5~GHz image superposed on a
0.33~GHz image of \fsrc.  In contrast to its triple morphology at
higher frequencies, \fsrc\ consists of a single component at~0.33~GHz,
approximately 20\arcsec\ in diameter.

The structure of \fsrc\ at~0.33~GHz is due, in part, to radio-wave
scattering.  The alternative, that it reflects \emph{only} the
intrinsic structure of the source, seems unlikely.  The high-frequency
spectrum of \fsrc, extrapolated to~0.33~GHz and combined with modest
free-free absorption, can account for the flux density of \fsrc\
(\S\ref{sec:spectrum}).  If the 0.33~GHz structure is a halo, then the
spectra of the individual components must flatten considerably or turn
over below~1~GHz.  While some spectral flattening is seen for USS
sources, typically this occurs at frequencies below~0.1~GHz
(\cite{rlmcs94}).  Also, the halo would have to have a spectral index
steeper than the individual components to avoid detection in our
observations (\S\ref{sec:observe}) and those of Lazio \&
Cordes~(1998a).

We cannot rule out the possibility that a halo contributes to the
0.33~GHz structure, however.  Some FR~II galaxies have low frequency
halos that surround the lobes (\cite{cb96}).  A halo enclosing
components~A and~C (separation 12\arcsec) would be largely resolved
out by our observations; such a halo could have also escaped detection
in the observations summarized in Lazio \& Cordes~(1998a), either
because the previous observations were less sensitive than those
reported here or because they did not have sufficient angular
resolution.  The uncertainty in the amount of free-free absorption
along this line of sight also allows for the existence of a halo.  We
require an optical depth $\tau \approx 0.3$ to reconcile the observed
and expected flux densities (\S\ref{sec:spectrum}).  An optical depth
only slightly larger, $\tau = 0.5$, for instance, would permit a halo
with a 0.33~GHz flux density $S_\nu \approx 0.1$~Jy to exist while
remaining consistent with the observed flux density.

Lazio \& Cordes~(1998a) considered the scattering properties of \fsrc\
assuming that the 0.33~GHz structure represented only the contribution
from \hbox{component~A}.  In fact all three components will contribute
to the 0.33~GHz structure.  The observed diameter of a source at a
particular frequency can be modelled as the quadrature sum of the
intrinsic diameter and the scattering diameter.  It is assumed
commonly that scattering affects the apparent diameter of only
relatively compact sources.  However, if the scattering diameter is
comparable to that of the intrinsic diameter, then the apparent
diameter of the source will reflect the contribution from scattering,
\emph{even if} the intrinsic structure of the source is not compact
(\cite{cc74}).

In solving for the scattering diameter of \fsrc, we shall obtain a
range of allowed values because of our limited information about its
intrinsic structure at~0.33~GHz.  The observed diameter of \fsrc\
at~0.33~GHz is 19\arcsec.  We first obtain an upper limit on the
scattering diameter by assuming that component~A, which dominates the
flux density of the source at high frequencies, continues to dominate
at~0.33~GHz.  Component~A has a high-frequency diameter (assumed to be
intrinsic) of approximately 2\arcsec.  Assuming that the intrinsic
diameter is frequency independent, we find a scattering diameter
$\theta_s \le 19\arcsec$.  We obtain an approximate lower limit on the
scattering diameter by taking the 12\arcsec\ separation between
components~A and~C to be characteristic of the 0.33~GHz intrinsic
diameter.  In this case $\theta_s \gtrsim 14\arcsec$.  The equivalent
range on the 1~GHz scattering diameter is 1\farcs5--2\arcsec.

While large, the scattering diameter of \fsrc\ is not unusually so,
nor is it unique toward the \hbox{GC}.  The nearby extragalactic
sources B1739$-$298 and B1741$-$312 ($\ell = 357\fdg87$,
$b=-1\arcdeg$) have similar lines of sight through the entire disk of
the Galaxy.  Their diameters indicate that they are heavily scattered,
though not affected by the hyperstrong region in front of \sgra\
(\cite{lc98b}).  Their scattering diameters, scaled to~0.33~GHz, are
roughly 6\arcsec, only a factor of 2--3 smaller than that inferred for
\fsrc.

The shape of the image at~0.33~GHz is also \emph{irregular}.  We
characterize the image shape as irregular as opposed to anisotropic
because the image displays a ``mottled'' or ``lumpy'' appearance
(Figure~\ref{fig:overlay}).  The intrinsic structure, as inferred from
the high frequency observations, contributes little to the irregular
appearance.  We have convolved our 1.5~GHz image
(Figure~\ref{fig:Lband}) with a circular gaussian scattering disk,
15\arcsec\ in diameter.  The resulting image is comparable in size to
that of our 0.33~GHz image, but is considerably smoother.

Our lack of knowledge about the low-frequency, intrinsic structure of
\fsrc\ limits the extent to which we can explain this irregular
appearance.  This appearance is probably the result of both the
intrinsic structure at~0.33~GHz and anisotropic scattering.
Observations at frequencies between~0.33 and~1.4~GHz would be useful
in separating these effects.  Unfortunately, existing observations in
this frequency range do not have sufficient angular resolution even to
resolve the source (e.g., the 0.84~GHz MOST survey had a beam of
$90\arcsec \times 45\arcsec$, \cite{g94}).

\subsection{Uniform vs.\ Patchy Scattering Region}\label{sec:patchy}

Using a likelihood analysis of the OH maser scattering diameters and
source counts of extragalactic sources toward the GC, Lazio \&
Cordes~(1998b) concluded that the characteristic separation between
\sgra\ and GC scattering region is 150~pc, that the scattering region
is centered on $(\ell,b) = (0\arcdeg, 0\arcdeg)$, and that its extent
in Galactic longitude is between $\pm 0\fdg5$ and $\pm 1\arcdeg$.  As
a consequence, \sgra\ and the heavily broadened OH masers are seen
through a hyperstrong scattering region that covers at least a portion
of the \hbox{GC}.  The implied scattering diameter at~0.33~GHz is
12\arcsec\ for a Galactic source and at least 600\arcsec\ for an
extragalactic source.  In this section we consider three possibilities
that would allow \fsrc\ to be both extragalactic and not heavily
scattered.

The first possibility is that the hyperstrong scattering region is a
relatively homogeneous region of sufficiently limited angular extent
that it does not cover \fsrc.  While the scattering region's extent in
Galactic latitude is poorly constrained (\cite{lc98b}), we now show
that if the scattering region is homogeneous, it almost certainly
extends to at least the Galactic latitude of \fsrc.  The maser
OH~359.762$+$0.120 is 7\farcm4 south of \fsrc\ and has an anisotropic
scattering disk of $1\farcs2 \times 0\farcs36$ at~1.6~GHz
(\cite{fdcv94}).  The nonthermal thread G0.08$+$0.15 is 13\arcmin\
northeast of \fsrc, and its width may be affected by scatter
broadening (\cite{apeg91}).  The nonthermal thread G359.79$+$0.17 is
4\farcm7 south of \fsrc\ (Figure~\ref{fig:Pband}), and as we now show,
its width is also likely to be affected by scatter broadening.  From a
1.4~GHz image provided by C.~C.~Lang (1998, private communication), we
have determined that the width of the thread varies from~12\arcsec\
to~19\arcsec; an example is shown in Figure~\ref{fig:thread}.  At this
frequency the expected scattering diameter for a Galactic source
affected by the hyperstrong scattering region is 0\farcs6, so we
consider these widths to be the intrinsic widths of the thread.  As
noted above, the expected scattering diameter for a Galactic source
at~0.33~GHz is 12\arcsec.  If we assume that the intrinsic width is
frequency independent and add the intrinsic width and scattering
diameter in quadrature, the expected width of the thread should vary
between~17\arcsec\ and~22\arcsec.  In fact, the observed width of the
thread at~0.33~GHz varies between 23\arcsec\ and~29\arcsec.

We regard the agreement between the expected and observed widths of
the thread as evidence that scattering affects the thread
G359.79$+$0.17.  The modest discrepancies between the expected and
observed width can be resolved by a combination of a
frequency-dependent intrinsic width and a larger separation between
the thread and scattering region than assumed.  A frequency-dependent
intrinsic width would arise if the synchrotron emitting electrons
responsible for the thread have a greater transverse extent at low
energies (\cite{apeg91}).  The expected scattering
diameter depends upon the distance between the source and the
scattering region.  We have taken the scattering diameter to be that
of \sgra, which is appropriate if the thread G359.79$+$0.17 is also
150~pc behind the scattering region (\cite{lc98b}).  If the thread is
more distant, its expected scattering diameter is
\begin{equation}
\theta_s 
 \simeq \theta_{\mathrm{Sgr~A^*}}\frac{\Delta}{\Delta_{\mathrm{Sgr~A^*}}},
\label{eqn:distance}
\end{equation}
where $\theta_{\mathrm{Sgr~A^*}}$ is the scattering diameter of \sgra,
$\Delta_{\mathrm{Sgr~A^*}}$ is the \sgra-scattering region separation,
and $\Delta$ is the thread-scattering region separation.  If the
thread G359.79$+$0.17 is 100~pc more distant than \sgra, scattering
alone could explain the discrepancy between the observed and expected
widths of the thread.

Therefore, if the scattering region is homogeneous and covers the GC
uniformly, it extends to a Galactic latitude of at least 0\fdg18.
Since \fsrc\ does not display the expected enhanced broadening, we
conclude that the scattering region is unlikely to be homogeneous.

The second possibility is that the scattering region is patchy or
clumpy and that there are ``holes'' in it.  If so, the size of the
clumps would have to be smaller than the 4\farcm7 separation between
\fsrc\ and the thread G359.79$+$0.17, equivalent to 12~pc at the
\hbox{GC}.

A scattering region composed of clumps would also explain the
contrasts between the diameters of OH~0.190$+$0.036 and
OH~0.319$-$0.040 and of Sgr~A OH1720:B and \hbox{Sgr~A OH1720:C}.  The
(1.6~GHz) scattering diameter of OH~0.190$+$0.036 is only 0\farcs1
(\cite{vd91}), whereas OH~0.319$-$0.040 has a scattering diameter
of~1\farcs2 (\cite{fdcv94}), even though it is only 9\arcmin\ away and
(angularly) more distant from \sgra.  The case of OH~0.190$+$0.036 is
not as clear-cut as that of \fsrc, however, because OH~0.190$+$0.036
could be just behind or in front of the scattering region.  While we
cannot establish definitely that OH~0.190$+$0.036 is close to the GC,
there are two indications that this is likely.  First, in order for it
to be unaffected by the scattering region, OH~0.190$+$0.036 would have
to be at least 100~pc closer to the Sun than \sgra\ (i.e., no more
than 50~pc behind the scattering region).  Its projected distance from
\sgra\ is 30~pc.  OH/IR stars in the GC have a volume density that
varies as $r^{-2}$ and a distribution that is centered on \sgra\
(Lindqvist, Habing, \& Winnberg~1992b).  Therefore, it is ten times
more likely for an OH/IR star to be 30~pc distant from \sgra\ than it
is to be 100~pc distant.  Second, OH/IR stars show a $\ell$-$v$
diagram indicative of circular rotation about the center
(\cite{lhw92}).  The radial velocity of OH~0.190$+$0.036 ($+159$~\kms,
\cite{lwhm92}) is not only consistent with that expected from circular
rotation, but is actually larger than the value predicted by Lindqvist
et al.'s~(1992b) model.  A radial velocity larger than the model value
is exactly what is expected if most of the the source's velocity is
projected along the line of sight, i.e., if OH~0.190$+$0.036 is at its
projected distance from \sgra.

Sgr~A OH1720:B is a collection of 1720~MHz OH maser features in Sgr~A
West having anisotropic angular diameters with major axes of
approximately 0\farcs85.  Sgr~A OH1720:C is an unresolved 1720~MHz OH
maser located 15\arcmin\ ($\approx 35$~pc) away and associated with
Sgr~A East (\cite{y-zrgfg98}).  In this case, Sgr~A West is known to
lie in front of Sgr~A East (\cite{paegvsz89}).  Thus, if the
scattering region covered the GC uniformly, Sgr~A OH1720:C would be
expected to show a large scattering disk.

A third possibility for the scattering of \fsrc\ is that the
scattering region is far from \sgra\ ($> 1$~kpc).  We do not consider
this option to be viable.  In their likelihood analysis of the
scattering region toward the GC, Lazio \& Cordes~(1998b) found that,
even if they assumed \fsrc\ to be extragalactic, there was still a
deficit of sources near \sgra, requiring the scattering region to be
closer than 500~pc to the \hbox{GC}.  (Their analysis also implicitly
assumed a relatively homogeneous scattering region.)  Moreover, their
likelihood analysis of the OH maser scattering diameters---which was
independent of any extragalactic sources---indicated that the
\sgra-scattering region separation is between~50 and~300~pc.
Nonetheless, the scattering diameter of \fsrc\ is formidable,
1\farcs5--2\arcsec\ at~1~GHz.  This diameter is comparable to that
seen for other heavily scattered sources in the Galaxy (e.g.,
1849$+$005, Fey, Spangler, \& Cordes~1991; NGC6634B, \cite{mrgb90}).
Since the mean free path for intersecting a comparable region of
scattering is 8~kpc (\cite{cwfsr91}), there is a good chance there
could be additional scattering contributed by a region on the far side
of the Galaxy.

\subsection{Implications for the Scattering Region}\label{sec:imply}

We therefore conclude that the scattering region is inhomogeneous or
clumpy on scales of 10--20~pc.  Prevailing models for the scattering
region explain the scattering as arising in the ionized outer layers
of molecular clouds, which are produced either from photoionization by
the hot stars in the GC (\cite{y-zcwmr94}) or as an evaporative
interface between the molecular clouds and the ambient X-ray gas
(\cite{lc98b}).  Inhomogeneities in the scattering region on scales
of~10--20~pc are consistent with a molecular cloud origin: Molecular
clouds in the GC also have scale sizes of~10--20~pc (\cite{rg89}).

We have examined CO ($J = 1 \to 0$, \cite{bswh87}; $J = 2 \to 1$,
\cite{ohhhs98}) $\ell$-$v$ diagrams of the GC in an effort to identify
a ``gap'' in the molecular cloud distribution through which \fsrc\
could shine.  We focus on CO because its emission traces densities $n
\gtrsim 500$~cm${}^{-3}$.  Since typical molecular cloud central
densities in the GC are $n \gtrsim 10^4$~cm${}^{-3}$, CO will trace
the outer layers of molecular clouds, the regions that can be ionized
easily.  In addition, high resolution, sensitive surveys exist for
these transitions, with resolutions comparable to that which we infer
for the separation between scattering clumps.  The $J = 1 \to 0$
observations had a 100\arcsec\ beam ($= 4$~pc at the distance of the
GC); the $J = 2 \to 1$ observations had a 9\arcmin\ beam ($= 20$~pc).

The distribution of molecular clouds in the GC displays a well-known
asymmetry with a concentration of molecular material toward $\ell >
0\arcdeg$.  This asymmetry means that a hole in the scattering region
for $\ell < 0\arcdeg$ is quite plausible.  Indeed, perhaps more
difficult to explain is the enhanced angular broadening toward the
thread G359.79$+$0.17 and OH~359.762$+$0.120, since there do not
appear to be obvious GC molecular clouds in these directions.  The
molecular cloud(s) (and the associated ionized outer layers)
responsible for the enhanced broadening toward the thread
G359.79$+$0.17 and OH~359.762$+$0.120 may have velocities near~0~\kms.
Most GC molecular clouds can be identified because they have large
velocities.  Some GC molecular clouds do have velocities near~0~\kms,
however, and these clouds can be confused with gas along the line of
sight to the \hbox{GC}.

We distinguish between the \emph{outer scale} in the scattering region
and the \emph{separation} between scattering clumps.  The outer scale,
$l_0$, is simply the largest scale on which the density fluctuations
responsible for radio-wave scattering occur.  If the density
fluctuations arise from a turbulent process, the outer scale reflects
the largest scale on which turbulent energy is injected.  If the
scattering arises in the ionized, outer layers of molecular clouds
(\cite{y-zcwmr94}; \cite{lc98b}), the outer scale presumably is comparable
the thickness of these ionized layers.  In both models, the outer
scale is small, $l_0 \lesssim 10^{-4}$~pc, but it is not necessarily
related to the separation between individual clouds.

The small outer scale (and high temperature \cite{lc98b}) of the
scattering region (or cloudlets) means that it contributes little to
the free-free emission and absorption toward the \hbox{GC}.  Rather
the bulk of the free-free emission and absorption would be congributed
by the extended, low-density \ion{H}{2} region (\cite{mp79};
\cite{ara97}).  This extended, low-density \ion{H}{2} region may be
(partially) responsible for the scattering disks of \fsrc,
B1739$-$298, and B1741$-$312.  The hyperstrong scattering seen toward
\sgra\ and various OH masers is due to this clumpy, GC scattering region.

\section{Conclusions}\label{sec:conclude}

We have presented new observations of the source \fsrc\ at frequencies
between~0.33 and~15~GHz and an \ion{H}{1} absorption spectrum.  We
conclude that this source is a Fanaroff-Riley~II radio galaxy, based
on the following characteristics:
\begin{description}
\item[Morphology] The source is a collinear triple source, with the
northeast component being edge brightened, the central component
having a jet-like feature pointing toward the NE component, and the
southwest component being extended toward the central component.
\item[Polarization] The edge-brightened component shows a fractional
polarization of approximately 10\%.
\item[Spectrum] Steep spectrum, with spectral index $\alpha < -1$,
characteristic of high-redshift galaxies.
\item[Variability] No significant structural or flux density
variations have been observed over short (few weeks to a month) or
long (10--15~yr) time scales.
\end{description}
Steep spectrum radio galaxies are typically at $z \gtrsim 2$.  We have
not been able to determine a distance to the source, but the
\ion{H}{1} absorption spectrum we have presented suggests a lower
limit of~25~kpc, and the classification as an FR~II radio galaxy
suggests a lower limit to the redshift of~0.1.

Though \fsrc\ is only 15\arcmin\ from \sgra, it is not affected by the
hyperstrong scattering region that is responsible for the large
diameter of \sgra.  If it were, \fsrc\ would have a diameter in excess
of 600\arcsec\ at~0.33~GHz in contrast to the observed diameter
of~20\arcsec.  We attribute the lack of hyperstrong scattering toward
\fsrc\ to an inhomogeneous scattering region.  An inhomogeneous
scattering region would also explain the small diameter of the masers
OH~0.190$+$0.036 and Sgr~A OH1720:C, both of which appear to be seen
through the region of enhanced scattering but have diameters less than
0\farcs1 at~1.6 and~1.7~GHz, respectively.  The scattering region
would have to be inhomogeneous on~5--10\arcmin\ scales (10--20~pc at
the Galactic center) to explain the small diameters of these two
sources.  This scale is comparable to the scale of molecular clouds in
the \hbox{GC}.  Prevailing models explain the hyperstrong scattering
as due to ionized regions on the surfaces of molecular clouds.  We
regard the close correspondence between the size of molecular clouds
and the inhomogeneities in the scattering region as evidence for a
connection between the two.

\acknowledgements

We thank F.~Yusef-Zadeh for stimulating discussions on the nature of
\fsrc, C.~Lang for providing the 1.4~GHz observations of the thread
359.79$+$0.17, C.~Carilli for helpful discussions on the properties of
FR~II radio galaxies, and L.~Sjouwerman and A.~Winnberg for comments
on the properties of OH masers toward the \hbox{GC}.  TJWL is
supported by a National Research Council-Naval Research Laboratory
Research Associateship.  Basic research in astronomy at the NRL is
supported by the Office of Naval Research.  The National Radio
Astronomy Observatory is a facility of the National Science Foundation
operated under cooperative agreement by Associated Universities, Inc.
The National Astronomy \& Ionosphere Center is operated by Cornell
University under contract with the \hbox{NSF}.  This research made use
of the SIMBAD database, operated at CDS, Strasbourg, France.

\clearpage

\begin{figure}
\caption[]{\fsrc\ and the nonthermal thread G359.79$+$0.17
at~0.33~GHz.  The off-source noise level in the image is 1.2~\mjybm,
and the contour levels are $-2.5$, 2.5, 3.54, 5, 7.07, 10,
14.1, 20, \ldots~\mjybm.  The beam is $9\farcs4 \times 5\farcs1$ with
a position angle of~$-2\arcdeg$.}
\label{fig:Pband}
\end{figure}

\begin{figure}
\caption[]{\fsrc\ at~1.5~GHz.  This image was formed by combining data
acquired at~1.42 and~1.66~GHz.  The off-source noise level in the
image is 0.59~\mjybm, and contours $-0.5$, 0.5, 0.707, 1, 1.41, 2,
\ldots~\mjybm.  The beam is shown in the lower left.}
\label{fig:Lband}
\end{figure}

\begin{figure}
\caption[]{\fsrc\ at~8.5~GHz.  
\textit{(a)}~The high-resolution image formed from the 1998
observations.  The off-source noise level in the image is
28~$\mu$Jy~beam${}^{-1}$, and contours are $-0.1$, 0.1, 0.141, 0.2,
0.283, 0.4, \ldots~\mjybm.  The beam is shown in the lower left.
\textit{(b)}~The lower-resolution image formed by combining all data
from observations in~1990 and~1998.  The off-source noise level in the
image is 40~$\mu$Jy~beam${}^{-1}$, and contours are $-0.05$, 0.05,
0.071, 0.1, 0.141, 0.2, \ldots~\mjybm.  The beam is shown in the lower
left.}
\label{fig:Xband}
\end{figure}

\begin{figure}
\caption[]{\fsrc\ at~15~GHz.  This image was formed by combining all
data from observations in 1998~April.  Component~C was not detected.
The off-source noise level in the image is 49~$\mu$Jy~beam${}^{-1}$;
contours are $-0.25$, 0.25, 0.354, 0.5, 0.707, 1, 1.41, 2, \ldots~\mjybm.  The beam is shown in the lower left.}
\label{fig:Uband}
\end{figure}

\begin{figure}
\caption[]{(a)~\ion{H}{1} absorption spectrum toward \sgra.
(b)~\ion{H}{1} absorption spectrum toward \fsrc.  Both spectra have a
velocity resolution of 2.6~\kms.}
\label{fig:hi}
\end{figure}

\begin{figure}
\caption[]{Spectra of components of \fsrc\ as derived from the
observations reported in this paper.  Crosses indicate flux density
measurements of component~A, filled squares are for component~B, and
stars are for \hbox{component~C}.  At~15~GHz only an upper limit can
be placed on the flux density of \hbox{component~C}.  Filled circles
show the combined flux density for all three components; at~0.33~GHz,
the individual components cannot be distinguished.}
\label{fig:spectra}
\end{figure}

\begin{figure}
\caption[]{A comparison of the structure of \fsrc\ at~0.33
and~8.5~GHz.  The 8.5~GHz contour levels are as in
Figure~\ref{fig:Xband}b; the 0.33~GHz 
contours are $-2$, 2, 4, 6, 8, 12, 16, 20, 28, 36, and~44~\mjybm.}
\label{fig:overlay}
\end{figure}

\begin{figure}
\caption[]{Two cross-cuts perpendicular to the major axis
of thread G359.79$+$0.17.  (a)~1.4~GHz.  (b)~0.33~GHz; this cut was
formed from the image shown in Figure~\ref{fig:Pband}.}
\label{fig:thread}
\end{figure}

\clearpage

\begin{deluxetable}{lccccc}
\tablecaption{VLA Observing Log for \fsrc\label{tab:observe}}
\tablehead{ & & \colhead{VLA} & \colhead{Observing} & & 
	\colhead{Image}
	\\
	\colhead{Frequency} & \colhead{Epoch} 
	& \colhead{Configuration} & \colhead{Time} 
	& \colhead{Beam} & \colhead{Noise Level} 
	\\
	\colhead{(GHz)} & & & & \colhead{(\arcsec)} & \colhead{(\mjybm)}}
\startdata
0.33 & 1996 October~19 & A & 8~hr & $9.4 \times 4.3$ & 1.7\phn\nl
1.5  & 1998 March~11   & A & 90~min. & $2.5 \times 1$ & 0.58\nl 
8.5  & 1990 October~22 & C & 30~min. & $5.7 \times 2.4$ & 0.11\nl
     & 1998 March~11   & A & 15~min. & $0.67 \times 0.23$ & 0.052 \nl
     & 1998 April~11   & A\tablenotemark{a}	& 90~min. & $0.53 \times 0.32$ & 0.032 \nl
     & all 1998 data   & A & 105~min. & $0.51 \times 0.30$ & 0.028 \nl
     & all data        & \nodata & 2.25~hr & $0.95 \times 0.67$ & 0.040\nl
15   & 1990 October 22 & C & 30~min. & $2.6 \times 1.2$ & \nodata \nl
     & 1998 April~11   & A\tablenotemark{a}	& 4.5~hr & $0.33 \times 0.20$ & \nodata \nl
     & 1998 April~16   & A & 100~min. & $0.32 \times 0.15$ & \nodata \nl
     & 1998 April~23   & A & 70~min. & $0.58 \times 0.13$ & \nodata \nl
     & all data        & \nodata & 7.3~hr & $0.38 \times 0.16$ & 0.049 \nl
\ion{H}{1} & 1991 October 5--6 & BnA & $2 \times 9$~hr & $4\farcs7 \times 3\farcs6$ & \nodata \nl

\enddata

\tablenotetext{a}{The VLA was in the A configuration, but only a
13-antenna subarray was used in the observations.}

\end{deluxetable}

\begin{deluxetable}{lccccc}
\tablecaption{Flux Density of \fsrc\
	Components\label{tab:fluxdensity}}
\tablehead{\colhead{Frequency} & \colhead{A} & \colhead{B} &
	\colhead{C} & \colhead{Total} \\
	\colhead{(GHz)} & \colhead{(mJy)} & \colhead{(mJy)} &
	\colhead{(mJy)} & \colhead{(mJy)}}

\startdata
0.33           & \nodata     & \nodata & \nodata  & 236\phd\phn \nl
1.55           & 141\phd\phn & 12.7\phn & 3.9\phn & 157\phd\phn \nl
8.5\phn        & \phn22.2    & \phn4.82 & 0.63    & \phn27.7 \nl
15\phd\phn\phn & \phn11.2    & \phn2.29 & $<$0.24 & \phn13.5 \nl

\enddata
\end{deluxetable}

\end{document}